\documentclass[aps,twocolumn,preprintnumbers,10pt,showpacs,superscriptaddress]{revtex4-1}
\usepackage[english]{babel}

\usepackage{dcolumn}
\usepackage{amsmath}
\usepackage{amsthm,amssymb}
\usepackage{amstext}
\usepackage{orcidlink}
\usepackage{graphicx,subfigure,setspace,soul}

\usepackage{hyperref}
\def\eps{\epsilon}
\newcommand{\dd}{\mathrm{d}}

\allowdisplaybreaks

\begin{document}

\title{An Analytic Computation of Three--Loop Five--Point Feynman Integrals}
\author{Yuanche Liu\,\orcidlink{0009-0008-4604-1306}}
\email{liuyuanche@mail.ustc.edu.cn}	
\affiliation{School of Physical Sciences, University of Science and Technology of China, Hefei, Anhui 230026, China}

\author{Antonela Matija\v si\'c}
\email{amatijas@uni-mainz.de}	
\affiliation{PRISMA Cluster of Excellence, Institut f\"ur Physik, Johannes Gutenberg-Universit\"at Mainz, Staudingerweg 7, 55099 Mainz, Germany}

\author{Julian Miczajka}
\email{julian.m@gmx.net}
\affiliation{Max-Planck-Institut f\"ur Physik, Werner-Heisenberg-Institut, 85748 Garching bei M\"unchen, Germany (at the time of the study)}

\author{Yingxuan Xu}
\email{yingxu@physik.hu-berlin.de}	
\affiliation{Humboldt--Universit\"at zu Berlin, Institut f\"ur Physik, Newtonstraße 15, 12489 Berlin, Germany}

\author{Yongqun Xu\,\orcidlink{0000-0001-7775-3498}}
\email{yongqunxu@mail.ustc.edu.cn}	
\affiliation{Interdisciplinary Center for Theoretical Study, University of Science and Technology of China, Hefei, Anhui 230026, China}

\author{Yang Zhang\,\orcidlink{0000-0001-9151-8486}}
\email{yzhphy@ustc.edu.cn}	
\affiliation{Interdisciplinary Center for Theoretical Study, University of Science and Technology of China, Hefei, Anhui 230026, China}
\affiliation{Peng Huanwu Center for Fundamental Theory, Hefei, Anhui 230026, China}


\preprint{HU-EP-24/38-RTG, MITP-24-085, MPP-2024-235, USTC-ICTS/PCFT-24-50} 
\date{\today}

\begin{abstract}
We evaluate the three--loop five--point pentagon-box-box massless integral family in the dimensional regularization scheme, via canonical differential equation. We use tools from computational algebraic geometry to enable the necessary integral reductions. The boundary values of the differential equation are determined analytically in the Euclidean region. 
To express the final result, we introduce a new representation of weight six functions in terms of one--fold integrals over the product of weight--three functions with weight--two kernels that are derived from the differential equation. Our work paves the way to the analytic computation of three--loop multi--leg Feynman integrals.

\end{abstract}
\maketitle
\section{Introduction}

High--precision predictions for scattering amplitudes are essential for data analysis at particle colliders. These predictions enable the exploration of fundamental questions, including the structure of the Standard Model and the search for new physics. The High-Luminosity Large Hadron Collider (HL-LHC) upgrade~\cite{ZurbanoFernandez:2020cco}, along with upcoming colliders such as the Future Circular Collider (FCC)~\cite{FCC:2024lyi}, International Linear Collider (ILC)~\cite{ILC:2007bjz}, and Circular Electron Positron Collider (CEPC)~\cite{CEPCStudyGroup:2023quu}, will achieve unprecedented experimental precision. This progress drives the need for advancements in theoretical methods.

To meet the demands of this accuracy, multi--loop perturbative computations are crucial for refining theoretical predictions~\cite{Begel:2022kwp,Huss:2022ful,Caola:2022ayt,Andersen:2024czj,ParticleDataGroup:2024cfk}. At the LHC energy scales, the strong coupling constant, $\alpha_s$, is the largest of the Standard Model coupling constants but remains small enough for perturbative calculations. Its uncertainty significantly affects the precision of theoretical predictions for cross--sections. Jet production observables, such as the ratio of two-- to three--jet cross sections, provide key opportunities to measure $\alpha_s$ with high precision. These measurements not only refine collider predictions but also enhance our understanding of potential new physics. Higher--order corrections in $\alpha_s$ for complex processes require more precise multi--loop evaluations, a task which becomes challenging with increasing loop orders and external legs.


One of the challenging parts of a multi--loop perturbative computation is the calculation of the appearing Feynman integrals. Significant progress has been made in this direction, with innovative techniques developed, such as differential equations~\cite{Henn:2013pwa}, sector decomposition~\cite{Borowka:2017idc}, auxiliary mass flow~\cite{Liu:2022chg,Liu:2017jxz}, difference equation~\cite{Lee:2012te}, etc.. In particular, many multi--loop Feynman integrals with four, five, and even six external legs, important for phenomenology and theory, were calculated analytically~\cite{Chicherin:2018old,Abreu:2018rcw,Chicherin:2020oor,Abreu:2020jxa,Canko:2020ylt,Abreu:2021smk,Chicherin:2021dyp,Abreu:2023rco,Abreu:2024yit,Jiang:2024eaj,Henn:2023vbd,DiVita:2014pza,Long:2024bmi,Henn:2021cyv,Henn:2024ngj,Chen:2022yni,Chen:2022pdw}. The analytic expressions, when coded in public packages, make the phenomenology computation highly efficient. Analytic results of Feynman integrals also provide the basis for theoretical studies in perturbative quantum field theory.



Despite significant advances, the analytic computation of multi--loop, multi--leg Feynman integrals in the dimensional regularization (DR) scheme remains a major challenge. Notably, in the literature, no analytic solutions exist for the three-loop five--point integral family in DR. Furthermore, for three--loop integrals with more than four external legs, even existing numerical methods struggle to provide efficient results. 


In this {\it Letter}, we present the first analytic evaluation of a three--loop five--point Feynman integral family. The solutions to the canonical differential equation (CDE) are expressed in terms of classical polylogarithms up to weight three, while the higher weight parts, up to weight six, are expressed as one--fold integrals. 




A major difficulty is the complexity of integration--by--parts (IBP) reduction. For typical three--loop integrals with more than four external legs, standard reduction tools ~\cite{Klappert:2020nbg, Smirnov:2019qkx} encounter great difficulties. We overcome this by using computational algebraic geometry--based IBP methods and the \verb|NeatIBP| package ~\cite{Wu:2023upw}, enabling the reduction of three--loop five--point integrals.

Another significant challenge is to find an integral basis with uniform transcendental (UT) weights. 
We apply the latest leading singularity analysis~\cite{Jiang:2024eaj,Chen:2022lzr}, the \verb|INITIAL| algorithm~\cite{Dlapa:2022wdu} and $\dd\log$ integrand construction~\cite{Henn:2020lye} to get a UT basis and thus the canonical differential equation.

In order to get the weight--six solution to the differential equation, we introduce a novel iterated integral representation. 
The solutions at weights four to six are all expressed as one--fold integrations over products of classical polylogarithms. Our solutions agree with standard numerical program \verb|pySecDec|~\cite{Borowka:2017idc,Li:2015foa,Borowka:2018goh} for all integrals, and with \verb|AMFlow|~\cite{Liu:2022chg,Liu:2017jxz} for the integrals calculable in the current version of \verb|AMFlow|.

Our approach marks a significant advancement in the analytic computation of three--loop multi--leg Feynman integrals, overcoming the constraints of traditional methods. We expect our result to play a crucial role in next--to--next--to--next--to--leading order (N$^3$LO) cross section computations for processes like three--jet production, three--photon production, jet--two--photon production, and the high--precision determination of $\alpha_s$. The result is also important for theoretical studies.

This {\it Letter} is organized as follows: In section \ref{sec:familydef}, we introduce the definition of the integral family. Section \ref{sec:IBP} explains how computational algebraic geometry methods enable the integral reduction. In section \ref{sec:ut}, we construct the UT basis, which is then used in section \ref{sec:de} to achieve the CDE. In section \ref{sec:func}, we express the integrals via simple functions at lower weight, which we supplement with the boundary values in section \ref{sec:BC}. In section \ref{sec:onefold}, a new representation of iterative integrals is used to express the higher weight integrals. Finally, section \ref{sec:con} provides concluding remarks and an outlook for future work. 



\section{Family Definition}\label{sec:familydef}
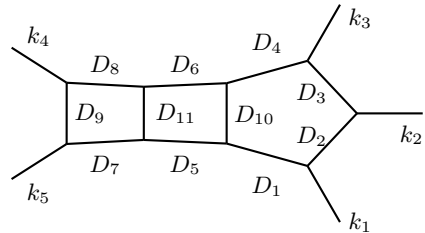
\begin{figure}[htbp]\centering
		\begin{tikzpicture}[thick, scale=.95]
			\draw  (4.14133,2.25835)--(3.01009,1.94511);
			\draw  (4.14133,2.25835)--(4.83319,1.52235);
			\draw  (4.14133,2.25835)--(4.5961,3.04481);
			\draw  (3.01009,1.94511)--(1.8514,1.89546);
			\draw  (3.01009,1.94511)--(3.00984,1.09978);
			\draw  (0.771529,1.95128)--(1.8514,1.89546);
			\draw  (0.771529,1.95128)--(0.77067,1.09397);
			\draw  (0.771529,1.95128)--(0.00103433,2.4426);
			\draw  (1.8514,1.89546)--(1.85057,1.1503);
			\draw  (3.00984,1.09978)--(4.14138,0.786459);
			\draw  (3.00984,1.09978)--(1.85057,1.1503);
			\draw  (4.14138,0.786459)--(4.83319,1.52235);
			\draw  (4.14138,0.786459)--(4.59549,0.);
			\draw  (1.85057,1.1503)--(0.77067,1.09397);
			\draw  (0.77067,1.09397)--(0.,0.602281);
			\draw  (4.83319,1.52235)--(5.75372,1.52222);
			\node[] at  (3.57561,0.5) {\small $D_1$};
			\node[] at  (3.57571,2.5) {\small $D_4$};
			\node[] at  (4.2,1.8) {\small $D_3$};
			\node[] at  (4.2,1.2) {\small $D_2$};
			\node[] at  (5.6,1.22) {\small $k_2$};
			\node[] at  (4.88,2.85158) {\small $k_3$};
			\node[] at  (4.88,0) {\small $k_1$};
			\node[] at  (3.4,1.5) {\small $D_{10}$};
			\node[] at  (1.1,1.52262) {\small $D_9$};
			\node[] at  (2.3,1.52288) {\small $D_{11}$};
			\node[] at  (2.4302,0.8) {\small $D_5$};
			\node[] at  (2.43074,2.2) {\small $D_6$};
			\node[] at  (1.31062,0.8) {\small $D_7$};
			\node[] at  (1.31146,2.2) {\small $D_8$};
			\node[] at  (0.38,0.4) {\small $k_5$};
			\node[] at  (0.38,2.6) {\small $k_4$};
		\end{tikzpicture}
		\caption{Pentagon-box-box Feynman Integral}\label{fig:2047}
\end{figure}

The three--loop five--point integral family  considered in this article is the so--called Pentagon--box--box (PBB) shown in Figure \ref{fig:2047}, while some sub--families are shown in Figure \ref{fig:sub}. All integrals from the family take the form
\begin{equation}
    I_{\nu_1,\dots, \nu_{18}}(\mathcal{N}) = e^{3\varepsilon \gamma_E}\int 
    \prod_{j=1}^3\frac{ \mathrm{d}^{4-2\varepsilon}\ell_j}{(i \pi)^{2-\varepsilon}} \frac{\mathcal{N}(\ell,k)}{\prod_{k=1}^{18}D_k^{\nu_k}}
    \label{eq:intdef}
\end{equation}
where $\varepsilon$ is the dimensional regulator, $\gamma_E$ is the Euler--Mascheroni constant, $\nu_k$ are the propagator powers and $\mathcal{N}$ is a kinematic numerator. The inverse propagators are defined as follows:
\begin{equation}
	\begin{array}{ll}
		&
		D_1=\ell _1^2,\quad
		D_2=\left(\ell _1-k_1\right){}^2,\notag\\
		&
		D_3=\left(\ell_1-k_1-k_2\right){}^2,
		D_4=\left(\ell _1-k_1-k_2-k_3\right){}^2,\notag \\
		&
		D_5=\ell_2^2,
		D_6=\left(\ell _2-k_1-k_2-k_3\right){}^2, \notag \\
		&
		D_7=\ell_3^2,
		D_8=\left(\ell_3-k_1-k_2-k_3\right){}^2,\notag\\
		&
		D_9=\left(\ell _3+k_5\right){}^2,
		D_{10}=\left(\ell_1-\ell_2\right){}^2,\notag\\
		&
		D_{11}=\left(\ell _2-\ell _3\right){}^2 ,
		D_{12}=\left(\ell_1+k_5\right){}^2,\notag\\
		&
		D_{13}=\left(\ell _2-k_1\right){}^2,
		D_{14}=\left(\ell_2-k_1-k_2\right){}^2,\notag\\
		&
		D_{15}=\left(\ell _2+k_5\right){}^2,
		D_{16}=\left(\ell_3-k_1\right){}^2, \\
		&
		D_{17}=\left(\ell _3-k_1-k_2\right){}^2,
		D_{18}=\left(\ell _1-\ell_3\right){}^2,
	\end{array}
\end{equation}
where $\{D_{i}|12\le i\le18\}$ serve as irreducible scalar products (ISPs) and $k_j$ are the external momenta of the scattering process. 

 We use standard kinematic variables, $s_{12}$, $s_{23}$, $s_{34}$, $s_{45}$, $s_{15}$, where $s_{ij}=2 k_i\cdot k_j$, and the parity-odd invariant
 \begin{equation}
	\eps_5=4i\epsilon_{\mu\nu\rho\sigma}k_1^\mu k_2^\nu k_3^\rho k_4^\sigma.
    \end{equation}

\section{Using Computational Algebraic Geometry for IBP Reduction}\label{sec:IBP}

IBP reductions are currently a crucial step in constructing a differential equation for an integral family. However, their computational complexity poses a bottleneck for most cutting--edge calculations both in terms of memory and computation time. To facilitate the computation of the PBB family, we thus turn to computational algebraic geometry methods~\cite{Gluza:2010ws,Georgoudis:2016wff,Badger:2013sta,Zhang:2012ce,Boehm:2020zig,Bohm:2018bdy,Larsen:2015ped,Wu:2024paw} and the public code \verb|NeatIBP|~\cite{Wu:2023upw}. \verb|NeatIBP| solves the module intersection problem to avoid the increase of propagator indices and gets a much shorter IBP linear system than that from the Laporta algorithm~\cite{Laporta:2000dsw}. 

On a laptop, \verb|NeatIBP| finds about $85000$ IBP identities in full kinematics dependence within several hours. These relations are sufficient to derive the differential equation. As a comparison, standard IBP tools would generate three orders of magnitude more IBP relations, which make the reduction computation impossible.

The PBB family as shown in Figure \ref{fig:2047} consists of $316$ master integrals, while details for some  sub--families are shown in Table \ref{tab:mis}.
\begin{figure}[htbp]
	\centering
 	\subfigure[Rocket]{
		\begin{minipage}[b]{0.13\textwidth}
					\begin{tikzpicture}[thick, scale=.33,cm={0,1,-1,0,(0,0)}]
							\draw  (0.273111,2.34496)--(2.27,2.30177);
							\draw  (0.273111,2.34496)--(0.270508,0.656094);
							\draw  (0.273111,2.34496)--(-1.,3.);
							\draw  (2.27,2.30177)--(4.41992,2.3511);
							\draw  (2.27,2.30177)--(2.27015,0.698757);
							\draw  (4.41955,0.647443)--(4.41992,2.3511);
							\draw  (4.41955,0.647443)--(2.27015,0.698757);
							\draw  (4.41955,0.647443)--(6.09807,1.5001);
							\draw  (4.41955,0.647443)--(4.3,-1.);
							\draw  (4.41992,2.3511)--(6.09807,1.5001);
							\draw  (4.41992,2.3511)--(4.3,4.);
							\draw  (2.27015,0.698757)--(0.270508,0.656094);
							\draw  (0.270508,0.656094)--(-1.,0.);
							\draw  (6.09807,1.5001)--(7.75,1.5);
            \node at  (1.5,0) {\footnotesize 7};
			\node at  (1.5,3) {\footnotesize 8};
			\node at  (0.7,1.5) {\footnotesize 9};
			\node at  (3.5,3) {\footnotesize 6};
			\node at  (3.5,0) {\footnotesize 5};
			\node at  (2.7,1.5) {\footnotesize 11};
			\node at  (4.0,1.5) {\footnotesize 10};
			\node at  (5.25,2.8) {\footnotesize 3};
			\node at  (5.25,0.2) {\footnotesize 2};
						\end{tikzpicture}
		\end{minipage}
		\label{fig:rocket}
	}
	\subfigure[Five--Point Ladder]{
		\begin{minipage}[b]{0.13\textwidth}
		\begin{tikzpicture}[thick, scale=.66]
				\draw (0, 0) rectangle (1, 1); 
				\draw (0, 1) rectangle (1, 2); 
				\draw (0, 2) rectangle (1, 3); 
				\draw (0, 0) -- (-0.6,-0.6);
				\draw (1, 0) -- (1.6,-0.6);
				\draw (1, 3) -- (1.6,3.6);
				\draw (0, 3) -- (-0.6,3.6);
				\draw (0, 2) -- (-0.6, 2);
                \node at (-0.5,2.5) {\footnotesize 2};
                \node at (0.5,3.25) {\footnotesize 3};
                \node at (1.5,2.5) {\footnotesize 4};
                \node at (-0.5,1.5) {\footnotesize 5};
                \node at (1.5,1.5) {\footnotesize 6};
                \node at (-0.5,0.5) {\footnotesize 7};
                \node at (1.5,0.5) {\footnotesize 8};
                \node at (0.5,1.25) {\footnotesize 11};
                \node at (0.5,0.25) {\footnotesize 9};
                \node at (0.5,2.25) {\footnotesize 10};
			\end{tikzpicture}
		\end{minipage}
		\label{fig:Five-Point-Ladder}
	}
 \subfigure[Box--Triangle--Triangle]{
		\begin{minipage}[b]{0.13\textwidth}
			\begin{tikzpicture}[thick, scale=.7]
				\draw (0, 1) -- (0.8, 2.2) -- (1.6,3);
				\draw (1, 1) -- (0.8, 2.2) -- (-0.1,2.5) -- (-.6, 3);
				\draw (-0.1, 2.5) -- (0, 1);
				\draw (-0.6,-0.8) -- (-0.1, -0.2) -- (0,1) --(1,1)-- (1.1,-0.2)-- (1.6,-0.8);
				\draw (-0.1,-0.2)--(1.1,-0.2);
				\draw (1,1)--(1.6,1);
                \node at  (0.5,-0.5) {\footnotesize 2};
                \node at  (0.5,0.6) {\footnotesize 10};
                \node at  (0.65,1.4) {\footnotesize 11};
                \node at  (0.5,2.6) {\footnotesize 9};
                \node at  (1.33,1.6) {\footnotesize 6};
                \node at  (1.5,0.5) {\footnotesize 3};
                \node at  (-0.35,0.5) {\footnotesize 1};
                \node at  (-0.5,1.77) {\footnotesize 7};
			\end{tikzpicture}
		\end{minipage}
		\label{fig:Box-Triangle-Triangle}}
	\caption{Typical Sub--Family for PBB}
	\label{fig:sub}\end{figure}
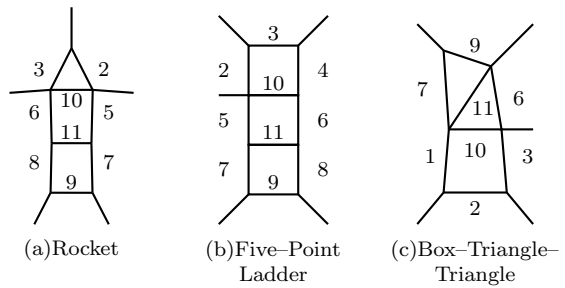
\begin{table}[htbp]\centering\begin{spacing}{1.25}
\begin{tabular}{ccc}\hline
    Family                  & Master Integrals     & Own Sector\\\hline
    The PBB                 &  316                 & 5  \\
    The Rocket              &  80                  & 6  \\
    Five--Point Ladder       &  175                 & 6  \\
    Box--Triangle--Triangle     &  49                  & 8  \\\hline
    \end{tabular}\caption{Typical Number of Master Integrals}\label{tab:mis}\end{spacing}
    \end{table}

We use the packages \verb|Rational Tracer|~\cite{Magerya:2022hvj} and \verb|FiniteFlow|~\cite{Peraro:2019svx} to solve the IBP system from \verb|NeatIBP| over finite fields. Once the UT basis and alphabet are determined, the canonical differential equation can be reconstructed from the alphabet. 
 
\section{UT Integral Determination}\label{sec:ut}
In this section, we use several computational techniques to complete the UT basis for a family of $316$ master integrals. 
Here is the basic strategy we took to search for the UT integrals: 
\begin{itemize}
	\item Classify the master integrals by the number of off--shell external legs resulting from pinching propagators. Among the $168$ topologies of the PBB family, $14$ correspond to two--point, $44$ to three--point, $80$ to four--point and $30$ for genuine five--point topologies. 
    
	\item The two-- and three--point UT topologies are quite straightforward. Moreover, from the works on three--loop four--point Feynman integrals with one off-shell leg~\cite{Henn:2023vbd,DiVita:2014pza}, we can find UT integrals for the four--point topologies.
	
	\item For the genuine five--point topologies with bubbles, we first integrate out the dotted bubble loop~\cite{Henn:2013tua} to get an effective two loop integral, and then look for the well studied two--loop five--point UT integrals~\cite{Chicherin:2018old}. For the remaining five--point topologies, we use \verb|Dlogbasis|~\cite{Henn:2020lye} algorithm  to search for dlog integrals. However, after this step, there are still some missing UT integrals.
	
	\item Those missing UT integrals are found case by case. One can make use of the \verb|INITIAL| algorithm, the Magnus series~\cite{Magnus:1954zz,Argeri:2014qva}, and some empirical rules to get the cut DE matrix in an $\varepsilon$--factorized form. If the off--diagonal parts are linear in $\varepsilon$, the $\mathcal{O}(\varepsilon^0)$ term can often be eliminated by subtracting tail integrals from lower sectors~\cite{He:2022ctv,Meyer:2016slj,Dlapa:2021qsl}.
\end{itemize}
To define the integrals in the novel genuine five--point sectors, we provide numerator factors that feature in \eqref{eq:intdef}.

\textbf{Pentagon--box--box} (Fig.\ref{fig:2047}): There are five UT integrals in the PBB sector given by
\begin{align}
	\mathcal{N}_1=&s_{12} s_{23} s_{45}^2D_{12}\label{int:322},\\
	\mathcal{N}_2=&s_{12} s_{23} s_{45}^2D_{15}-s_{23}  s_{45}^2D_1-s_{12} s_{45}^2D_4\label{int:-4},\\
	\mathcal{N}_3=&\frac{s_{45}^2}{\epsilon_5}G(\ell_1,k_1,k_2,k_3,k_4),\\
	\mathcal{N}_4=&\frac{s_{45}^2}{\epsilon_5}G\begin{pmatrix}
		\ell_1,k_1,k_2,k_3,k_4\\
		\ell_2,k_1,k_2,k_3,k_4\\
	\end{pmatrix},\\
	\mathcal{N}_5=&\frac{s_{45}^2}{\epsilon_5}G\begin{pmatrix}
		\ell_1,k_1,k_2,k_3,k_4\\
		\ell_3,k_1,k_2,k_3,k_4\\
	\end{pmatrix},
\end{align}
 where $G$ denotes the Gram determinant. Note that for the integral \eqref{int:-4}, we need to introduce a correction from a lower sector to make it a UT integral. 

\textbf{Five--Point Ladder} (Fig.\ref{fig:Five-Point-Ladder}): There are six UT integrals given by 
\begin{align}
	\mathcal{N}_1=&s_{23} s_{34} s_{45}^2,\\
	\mathcal{N}_2=&s_{23} s_{45}^2D_{12},\\
	\mathcal{N}_3=&s_{23} s_{45}^2D_{14},\\
	\mathcal{N}_4=&s_{23} s_{45} (s_{45}-s_{12})D_{15},\\
	\mathcal{N}_5=&\frac{s_{45}}{\epsilon_5}G\begin{pmatrix}
		\ell_1,k_1,k_2,k_3,k_4\\
		\ell_2,k_1,k_2,k_3,k_4\\
	\end{pmatrix},\\
	\mathcal{N}_6=&\frac{s_{45}}{\epsilon_5}G\begin{pmatrix}
		\ell_1,k_1,k_2,k_3,k_4\\
		\ell_3,k_1,k_2,k_3,k_4\\
	\end{pmatrix}.
\end{align}

\textbf{The Rocket} (Fig.\ref{fig:rocket}): Out of the six UT integrals in this sector, only five are constructed by \verb|Dlogbasis|. The last one can be found as follows. First, we observe that the DE matrix on cut is linear to $\varepsilon$, so the $\mathcal{O}(\varepsilon^0)$-term can be eliminated via Magnus series. After that, we turn to the full DE matrix and eliminate the $\mathcal{O}(\varepsilon^0)$-term by adding tail integrals. 

\textbf{Box--triangle--triangle} (Fig.\ref{fig:Box-Triangle-Triangle}):
We need eight UT integrals in this sector, while \verb|Dlogbasis| returns only three candidates. If we double some propagators on triangle~\cite{Henn:2013tua} and manually adjust an ansatz, it is not difficult to get the cut DE matrix in $\varepsilon$-form. After that, the off diagonal parts $\mathcal{O}(\varepsilon^0)$-terms can be eliminated by adding tail integrals. 

The full UT basis for the PPB family can be found in the auxiliary file \verb|Canonical_Basis.m|, see Appendix~\ref{sec:supp}.

\section{Differential Equation and Symbol Alphabet}\label{sec:de}

The pentagon alphabet has been extensively studied at two--loop order~\cite{Chicherin:2017dob}, and it was suggested that additional letters would appear at three--loop order~\cite{Chicherin:2024hes}. However, for the PBB family, the $31$ pentagon letters listed in Ref.~\cite{Chicherin:2017dob} are sufficient. Specifically, for the PBB family, only the planar pentagon letters $\mathbb{A}_{\text{P}}$ appear, while the non--planar letters $W_{20+i}, (1\le i\le5)$ are absent. 
The corresponding CDE can be decomposed as 
\begin{align}\label{eq:cderef}
	\mathrm{d}\mathbf{I}=\varepsilon \mathrm{d}\tilde{A} \mathbf{I}, \ \ \mathrm{d}\tilde{A}=\sum_{i=1}^{31}a_i \mathrm{dlog}W_i, 
\end{align}
where $W_i$ are algebraic functions of the kinematic variables and the $a_i$ are $\mathbb{Q}$--valued constant $316 \times 316$ matrices. 
The knowledge of the CDE allows us to derive an $\varepsilon$--expansion for any integral in the family 
\begin{equation}
    \mathbf{I} = \sum_{j=-6}^0 \varepsilon^{j} \mathbf{I}^{(j+6)} + \mathcal{O}(\varepsilon),
\end{equation}
where the coefficients $\mathbf{I}^{(n)}$ are referred to as the weight--$n$ part of the solution. In this letter, we are interested in the functions up to weight six that are relevant for applications to three--loop scattering processes.

\section{Pentagon Functions}\label{sec:func}

The functions at a given weight are most conveniently expressed in terms of a basis of transcendental functions. At five points, they are referred to as pentagon functions. They have been extensively studied at two--loop order ~\cite{Gehrmann:2015bfy,Chicherin:2017dob,Gehrmann:2018yef,Chicherin:2020oor,Chicherin:2021dyp,Abreu:2020jxa,Canko:2020ylt,Abreu:2021smk,Chicherin:2021dyp,Abreu:2023rco,Abreu:2024yit,Jiang:2024eaj,Caron-Huot:2020vlo}.  

For the PBB family, the weight--one and weight--two functions are simply 
\begin{equation}\label{eq:w1w2fun}
	\log(-W_i),\ \ \ \text{Li}_2\frac{W_{10+i}}{W_i}, \qquad 1\le i\le 5.
\end{equation}
These functions are well defined in the full Euclidean region $s_{i,1+(i)_5}<0$, $(1\le i \le 5)$. They are exactly the same as the functions that appear in the two--loop pentagon box integrals~\cite{Gehrmann:2018yef}.

At weight--three, the structure of the functions becomes more intricate. To express the integrals, logarithms with additional arguments are needed,
\begin{equation}
	\log \left(-W_{15+i}\right),\qquad 1\le i\le 5,\label{eq:extraw1}
\end{equation}
as well as $\text{Li}_2$ and $\text{Li}_3$ functions with the following arguments 
\begin{align}\label{eq:w2arg}
	&\text{Li}_n\left(   -\frac{W_{11 + (1 + i)_5}}{W_{i}}   		\right),
	\text{Li}_n\left(	\frac{W_{11 + (2 + i)_5}}{W_{i}}  	 		\right),\\&
	\text{Li}_n\left(	-\frac{W_{11 + (3 + i)_5}}{W_{i}}  			\right),
	\text{Li}_n\left(	-\frac{W_{11 + (2 + i)_5}}{W_{16+(4+i)_5}}	\right),\notag\\&
	\text{Li}_n\left(	\frac{W_{11 + (4 + i)_5}}{ W_{16+(4+i)_5} }	\right),
	\text{Li}_n\left(	-\frac{W_{11 + (2 + i)_5}W_{11 + (4 + i)_5}}{W_{i}W_{1+(1+i)_5}}	\right),\notag\\&
	\text{Li}_n\left(-\frac{W_{11 + (1 + i)_5}W_{11 + (4 + i)_5}}{W_{i}W_{16+(1+i)_5}}\right),\  1\le i\le 5, \ n\in \{2,3\}.\notag
\end{align}	
We use a subscript--5 notation, $(x)_5 \equiv \bmod(x,5)$, to denote that the pentagon functions form a group with $S_5$ cyclic permutation symmetry. To obtain such functions, we used the algorithm implemented in~\cite{Lee:2024kkm} to construct parity--even arguments for the $\text{Li}_n $ function.

The space formed by these functions is large enough to describe the entire family at weight--three except for one parity--odd integral, which is discussed in the following subsection.


\subsection{Matching the Parity--Odd Weight--Three Symbol}\label{sec:sbodd}


In the Euclidean region, the square root first appears at weight three. Initially, several integrals seem to involve parity--odd components at this weight. However, the choice of a UT basis for an integral family is not unique. It is possible to introduce $\mathbb{Q}$-linear combinations for these UT integrals. 
Therefore, the parity--odd components at weight three share one single parity--odd function, which we denote as $\mathcal{P}(s_{12}, s_{23}, s_{34}, s_{45}, s_{15})$.

However, even with this simplification, performing the integration remains challenging. Naive rationalization of the square root leads to a cumbersome expression with a complex branch cut structure, making it difficult to manipulate. 

Our approach is to first construct a function basis and then fit the symbol $\text{SB}(\mathcal{P})$ against it. At weight three, only classical polylogarithms are involved, and their symbol structure indicates that we shall search for $\text{Li}_n(x)$ such that both $x$ and $1-x$ factorized over the symbol alphabet~\cite{Duhr:2011zq,Heller:2019gkq,ediss28684}. This approach motivates a search for solutions over $\mathbb{Q}$ for the coefficients $q_i$ that satisfy
{\small \begin{align}\label{eq:dlog1-x}
		\mathrm{dlog}(1-x)=\sum_{i=1}^{31} q_i \mathrm{dlog} W_i, \ \mathrm{dlog}(x)=\sum_{i=1}^{31} q_i \mathrm{dlog} W_i.
	\end{align}}

We use a heuristic algorithm to find all necessary arguments for matching the symbol $\text{SB}(\mathcal{P})$. First, we perform the integration with rationalized square root via \verb*|PolyLogTools|~\cite{Duhr:2019tlz} and express the function $\mathcal{P}$ as multiple polylogarithms (MPLs). Next, we reduce the MPLs in terms of classical polylogarithms~\cite{Frellesvig:2016ske,DelDuca:2009ac} and extract the arguments which respect the condition \eqref{eq:dlog1-x}. We eventually reach identities like
{\small\begin{align}
		&2\text{dlog}\left(\frac{W_2 W_3 (W_{28}-1)}{W_{31}W_{28}}\right)=\notag\\
	&-\text{dlog}W_1+\text{dlog}W_2+\text{dlog}W_3-\text{dlog}W_{19}-
    \text{dlog}W_{28},\notag\\
	&2\text{dlog}\left(1-\frac{W_2 W_3 (W_{28}-1)}{W_{31}W_{28}}\right)=-\text{dlog}W_{1}+\text{dlog}W_{4}\notag\\
	&-\text{dlog}W_{19}+\text{dlog}W_{20}-\text{dlog}W_{28}-\text{dlog}W_{29}.
\end{align}}

With these arguments and the determined even part, we can find a function $\tilde{\mathcal{P}}$
{\begin{align}\label{eq:sbterm}
	\tilde{\mathcal{P}}&=
	...+2\text{Li}_3\left(\frac{W_2 W_3 (W_{28}-1)}{W_{31}W_{28}}\right)\notag\\
	&+2\text{Li}_3\left(\frac{\frac{2 W_{28} W_{31}}{W_{28}-1}-2 W_2 \left(W_1+W_5\right)}{\frac{2 W_{27} W_{31}}{W_{27}-1}-2 W_1 \left(W_2+W_3\right)}\right)+...\\
	&-2\text{Li}_2\left(\frac{W_{14} \left(W_1 W_2-\frac{W_{27} W_{31}}{W_{27}-1}\right)}{W_1 W_2 W_{19}}\right)\log\left(-W_1\right)+...,\notag
\end{align}}
such that $\text{SB}(\tilde{\mathcal{P}})=\text{SB}(\mathcal{P})$. Note that although there are only $40$ terms in $\text{SB}(\mathcal{P})$,  more than $2800$ weight--three functions have to be combined in order to fit the $\text{SB}(\mathcal{P})$. Also, we have to restrict the computation to a smaller region, where the arguments have definitive signs. To define $\mathcal{\tilde{P}}$ in other regions, one has to construct a basis for such a region or perform analytic continuation. 

We determine $\tilde{\mathcal{P}}$ by introducing $10$ additional $\log$ arguments, namely $\log W_i, $ with $\{11\le i\le15\}\cup \{16\le i\le20\}$, as well as $14$ new arguments for $\text{Li}_2$, and $23$ arguments for $\text{Li}_3$. Furthermore, the ``beyond--the--symbol" term $P-\tilde P$ can be found by numerically fitting the integrand after subtracting  $\tilde{\mathcal{P}}$ onto a basis of $\pi^2\log$. 

Finally, following these steps, we reach the function representation for the odd integral $\mathcal{P}$: 
\begin{align}
	\mathcal{P}=&\underbrace{\tilde{\mathcal{P}}}_{526\text{\ terms}}-\frac{\pi^2}{6}\Big(			
	 \log \left(-W_1\right)
	-2 \log \left(-W_2\right)
	+ \log \left(-W_3\right)
	\notag\\
	&+3 \log \left(-W_4\right)-2 \log \left(-W_{16}\right)
	-\log \left(-W_{20}\right)\notag\\&
	+ \log \left(-W_{26}\right)
	+ \log \left(-W_{27}\right)
	+ \log \left(-W_{28}\right)
	\notag\\&
	+2 \log \left(-W_{29}\right)
	-\log \left(-W_{30}\right)
	\Big)+ \zeta_3.
\end{align}
The constant $\zeta_3$ is found by the method described in the following section. At this point, we have already expressed all pentagon functions and thus all integrals in PBB family, up to weight--three, in terms of classical polylogarithms. 

\section{Boundary Values}\label{sec:BC}

The boundary values of the differential equation can be determined from the cancellation of spurious poles~\cite{Henn:2014lfa,Gehrmann:2018yef,Chicherin:2018mue,Henn:2024ngj}. To illustrate this idea, we adopt the following integral representation: 
\begin{equation}\label{eq:intrep}
	\mathbf{I}^{(n)}(x)=\underbrace{\mathbf{I}^{(n)}(x_0)}_{\text{Boundary\ Values}}+\underbrace{\int_{x_0}^x \dd \tilde A\cdot \mathbf{I}^{(n-1)}(x)}_{\text{Function Part\ } }.
\end{equation}
We denote the position of the spurious poles as $x_i$. If we require that the weight--$(n+1)$ integral should be finite at the spurious poles, then the integrand should vanish at these points:
\begin{equation}
        \text{Res}\bigg(\frac{\dd \tilde{A}}{\dd x}\bigg)\bigg|_{x_i}\cdot \mathbf{I}^{(n)}(x_i)=0.
\end{equation}
We require that the integrals are finite at these spurious poles $x_0=\{-1,-1,-1,-1,-1\}$, which is also chosen as the boundary point, $x_1=\{-2,-1,-1,-1,-1\}$ along with its cyclic permutations, and $x_6=\{1/4,-1,-1,-1,-1\}$, where the $\eps_5$ vanished. This allows us to express all boundary constants, at all orders in $\varepsilon$, as a $\mathbb{Q}$--combination of the transcendental functions that appear in the second term of \eqref{eq:intrep} evaluated at these spurious poles, up to one trivial integral
\begin{small}\begin{align}
		I_1(x_0)&=-1+\frac{\pi ^2 }{4}\varepsilon ^2+7 \zeta_3\varepsilon ^3+\frac{37 \pi ^4 }{480}\varepsilon ^4+\left(\frac{93 \zeta_5}{5}-\frac{7 \pi ^2 \zeta_3}{4}\right)\varepsilon ^5\notag\\&
		+\left(\frac{943 \pi ^6}{120960}-\frac{49 \zeta_3^2}{2}\right)\varepsilon ^6 +O\left(\varepsilon ^7\right).	
\end{align}\end{small}

Intriguingly, we found that the weight--three boundary constant for the integral $\mathcal{P}$ is numerically the same as the one in the one-- and two--loop cases~\cite{Gehrmann:2018yef}. This constant, originally written in terms of hundreds of MPLs~\cite{Gehrmann:2018yef}, can be analytically simplified to only four terms:
\begin{align}
	\mathcal{P}(-1,-1,-1,-1,-1)=&\notag\\-14 \text{Li}_3(-\phi )-\frac{14}{3}  \log ^3 &\phi-\frac{46}{15} \pi ^2 \log\phi+\frac{7}{5}\zeta_3,
\end{align}
where $\phi=\left(1+\sqrt{5}\right)/2$ is the golden ratio. 

Employing the above strategy, we analytically compute the boundary values up to weight--six for the entire PBB family. The package \verb|NumPolyLog|~\cite{Gint} and \verb|PSLQ| algorithm have been used to numerically evaluate and simplify the boundary values. 

\section{Auxiliary Matrix for Higher Weight integration}\label{sec:onefold}


With the analytic weight three solutions and the boundary constants at hand, we can turn towards the computation of the integrals up to weight six. However, beyond weight three, classical polylogarithms are no longer sufficient and the number and complexity of the functions grow significantly. For multi--scale problems with square roots, obtaining higher weight expressions in terms of MPLs is still a highly non--trivial task. Moreover, such explicit MPL representations are not always good for efficient numerical evaluations. Therefore, it is more practical to express the higher weight iterative integrals in terms of a one--fold integral representation~\cite{Caron-Huot:2014lda,Gehrmann:2018yef,Chicherin:2021dyp,Wang:2022enl,Henn:2024ngj}. 

Starting from the iterative representation of the solution of CDE 
\begin{equation}
    \mathbf{I}^{(n+1)}(x)= \mathbf{I}^{(n+1)}(x_0)+\int_0^1\frac{\mathrm{d}\tilde{A}(t)}{\mathrm{d t}}\mathbf{I}^{(n)}(t)\mathrm{d} t,
\end{equation}
we can reach the one--fold integration formula, which is widely used at two--loop order to reach weight--four solution from weight--two functions
\begin{align}
	\mathbf{I}^{(n+2)}(x)= &\mathbf{I}^{(n+2)}(x_0)
	+\int_0^1\frac{\mathrm{d}\tilde{A}(t)}{\mathrm{d t}}\mathbf{I}^{(n+1)}(x_0)\mathrm{d}t
	\notag\\&
	+\int_0^1(\tilde{A}(1)-\tilde{A}(t))\frac{\mathrm{d}\tilde{A}(t)}{\mathrm{d t}}\mathbf{I}^{(n)}(t)	\mathrm{d} t.\label{eq:onefold1}
\end{align}
If we wanted to apply this technique to calculate the weight six part of the PBB family, we would require analytic knowledge of the weight four function. To circumvent this requirement, we introduce a novel formula to write higher weight parts as one--fold integrals.
By defining a weight--two \textit{auxiliary matrix} $\tilde{B}$ as the integration kernel, which satisfies a matrix differential equation
\begin{equation}\label{eq:defb}
	\dd \tilde{B}=(\dd\tilde{A})\tilde{A},
\end{equation}
one can decompose a weight--$(n+3)$ function into a one--fold integration, 
{\small\begin{align}\label{eq:w6onefold}
	\mathbf{I}^{(n+3)}(x)
	=&\mathbf{I}^{(n+3)}(x_0)
	+\int_0^1\frac{\dd\tilde{A}}{\dd t}\mathbf{I}^{(n+2)}(x_0)\dd t
	\notag\\&
	+\int_0^1\left(\tilde{A}(1)-\tilde{A}(t)\right)
	\frac{\dd\tilde{A}}{\dd t}\mathbf{I}^{(n+1)}(x_0)\dd t\notag\\
	&+\int_0^1
	\left(
	\tilde{A}(t)-\tilde{A}(1)
	\right)
	\tilde{A}(t)
	\frac{\dd\tilde{A}}{\dd t}
	\mathbf{I}^{(n)}(t)
	\dd t
	\notag\\&
	+\int_0^1
	\left(\tilde{B}(1)-\tilde{B}(t)\right)
	\frac{\dd\tilde{A}}{\dd t}
	\mathbf{I}^{(n)}(t)
	\dd t.
\end{align}}
$\tilde B$ is guaranteed to exist because of the integrability condition  
\begin{equation}
\dd\bigg((\dd\tilde A) \tilde A\bigg) =-\frac{1}{2}[A_i,A_j] dx^i \wedge dx^j =0,
 \end{equation}
from the property of canonical differential equation.

In our case, we use this formula with $n=3$ to express the weight--six part in terms of one fold integration. Based on the same symbol matching techniques as in section \ref{sec:sbodd}, it is straightforward to express the $\tilde{B}$--kernel in terms of $\log$ and $\text{Li}_2$.

 
Alternatively, one can express the weight--six part as a one--fold integration over the weight--two integrals and new auxiliary matrix $\tilde{C}$ defined as $\dd \tilde{C}=(\dd \tilde{A}) \tilde{B}$. 
 $\tilde{C}$ matrix can be used to calculate the weight--$n$ part from analytic expressions of weight--$(n-4)$ solutions. 


 
\section{Conclusion and Outlook}\label{sec:con}

In this \emph{Letter}, we initiate the analytic computation of three--loop five--point Feynman integrals in dimensional regularization. The pentagon--box--box integral family is calculated up to weight six via the canonical differential equation method. Powered by a novel iterated integral formula, we achieve the explicit function representation up to weight three and the one--fold integration representation up to weight six in the Euclidean region, with only $\log$, $\text{Li}_2$ and $\text{Li}_3$ functions.  The boundary values for this integral family up to weight six are analytically expressed as special values of MPLs. A proof--of--concept implementation of one--fold integration is provided to calculate this integral family with high efficiency, see Appendix~\ref{sec:supp}.



Our results initiate the computation of the Feynman integrals for massless $2\to3$ processes at N$^{3}$LO, and thereby open a new level of precision for collider phenomenology. With the developments in the IBP reduction and iterative integrals, we expect the full computation of all planar three--loop five--point integrals to finish in the near future. 

We anticipate that the methods and the results we have presented will have numerous applications, for example to $3$--jet and $3$--photon production, and to the precision determination of the strong coupling constant. Moreover, together with the recent computation of two--loop six--point Feynman integrals~\cite{Henn:2021cyv,Henn:2024ngj}, the massless $2\to 3$ N${}^3$LO computations are within reach.

\section{Acknowledgment}
We acknowledge Dmitry Chicherin, Johannes Henn and Shunqing Zhang for discussion on the alphabet, Roman Lee for the help with the iterative integrals. We further thank Jiaqi Chen, Xuan Chen, John Ellis, Lihong Huang, David Kosower, Zhenjie Li, Yanqing Ma, Vitaly Magerya, Yu Wu, Zihao Wu, Huaxing Zhu, Simone Zoia for enlightening discussions.  Yuanche Liu is supported from the NSF of China through Grant No. 124B1014, Yingxuan Xu is funded by the Deutsche Forschungsgemeinschaft (DFG, German Research Foundation) – Projektnummer 417533893/GRK2575 “Rethinking Quantum Field Theory”. Yang Zhang is supported from the NSF of China through Grant No. 12075234, 12247103, and 12047502 and  thanks the Galileo Galilei Institute for Theoretical Physics for the hospitality and the INFN for partial support during the completion of this work. This research was supported by the Munich Institute for Astro-, Particle and BioPhysics (MIAPbP) which is funded by the Deutsche Forschungsgemeinschaft (DFG, German Research Foundation) under Germany´s Excellence Strategy – EXC-2094 – 390783311.

\appendix

\bibliographystyle{h-physrev}
\bibliography{ref.bib}

\section{Supplemental Material}\label{sec:supp}

In this appendix, we first briefly introduce some of the auxiliary files mentioned in this paper. Then we show how to validate our computation by the numerical cross check. 

    \begin{itemize}
            \item \verb|Canonical_Basis.m|, the UT basis for the PBB family. 
		\item \verb|Atilde.m|, the CDE matrix $\tilde{A}$ defined in Eq.\eqref{eq:cderef}
            \item \verb|Weight-<i>-Integrals.m|, analytic functions for the UT integrals from weight--one to weight--three. 
            \item \verb|Weight-<i>-Symbol.tar.xz|, compress file for the symbol of the UT integrals. 
            \item \verb|Boundary-Values-Analytics.m|, analytic boundary values from weight--one to weight--six.
		\item \verb|Btilde.m|, the $\tilde{B}$ kernel matrix define in Eq.\eqref{eq:defb}
		\item \verb|Evaluate-All-Integrals.wl|, a proof--of--concept implementation of the one-fold integration to numerically evaluate the UT integrals.
	\end{itemize}
    These auxiliary files can be accessed at:
\begin{center}
    \url{https://github.com/YongqunXu/3L5P_PBB}.
\end{center}
For a generic point in the Euclidean region 
\begin{equation}
\left\{-\frac{79}{46}, -\frac{129}{29}, -\frac{89}{36}, -\frac{87}{55}, -\frac{51}{13}\right\}
\end{equation}
 our proof--of--concept implementation for one--fold integration is sufficient to produce the numeric results with 32 digits for all UT integrals in less than one hour with $30$ cores. For comparison, it takes around 38 hours for \verb|pySecDec| with four \verb|A100| GPUs to generate a result for a single UT integral \eqref{int:322} in the top sector with several digits, which is shown in Table \ref{tab:pysecdec}.  Evaluating the full family with the current version of \verb|AMFlow| is not yet feasible. 

    \begin{table}[ht]\vspace{0.34em}\begin{spacing}{1.2}
        \begin{tabular}{ccc}
        \hline
		$\varepsilon$-order       & \verb|pySecDec| results&Analytic results \\      \hline
		$\varepsilon^0$& $+1.1666667\pm3.80\times 10^{-8}$  & $+1.1666666$  \\
		$\varepsilon^1$& $-3.2652194\pm1.20\times 10^{-6}$  & $-3.2652203$ \\
		$\varepsilon^2$& $-3.2525854\pm9.30\times 10^{-6}$  & $-3.2525818$ \\
		$\varepsilon^3$& $-3.3833398\pm4.82\times 10^{-5}$  & $-3.3833422$\\
		$\varepsilon^4$& $-37.8523289\pm1.96\times 10^{-4}$ & $-37.8523988$\\
		$\varepsilon^5$& $-122.2098436\pm7.81\times 10^{-4}$& $-122.2104218$\\
		$\varepsilon^6$& $-71.5043735\pm6.72\times 10^{-3}$ & $-71.5031577$\\		
        \hline
        \end{tabular}\end{spacing}\caption{Numerical Cross Check}\label{tab:pysecdec}
        \end{table}

\end{document}